\documentclass{article}
\usepackage{isolatin1}
\usepackage{indentfirst,amsmath,amssymb,mathrsfs}
\usepackage{bbold,fancybox}

\begin{document}

\newtheorem{theorem}{Theorem}[section]
\newtheorem{lemma}[theorem]{Lemma}
\newtheorem{proposition}[theorem]{Proposition}
\newtheorem{corollary}[theorem]{Corollary}
\newtheorem{definition}{Definition}[section]
\newtheorem{remark}[definition]{Remark}
\newtheorem{conjecture}{Conjecture}[section]

\title{Algebraic holography in asymptotically simple, asymptotically AdS 
        spacetimes \\{\small This work is dedicated 
	to Professor Jacques Bros, on the occasion of his 70th birthday.}}
\author{Pedro Lauridsen Ribeiro \thanks{Talk given during the Symposium ``Rigorous 
        Quantum Field Theory'' in the honour of Prof. J. Bros, SPhT-Saclay, Paris,
	July 19-21, 2004. This project is supported by 
	FAPESP under grant no. 01/14360-1.}\\
        Departamento de Física Matemática \\ Instituto de Física, 
	Universidade de São Paulo \\ CP 66.318 \quad 05315-970 \quad 
	São Paulo, SP -- Brasil \\ Email: \texttt{pribeiro@fma.if.usp.br}}
\maketitle

\begin{abstract}
\indent I'll describe a general geometric setup allowing 
a generalization of Rehren duality to asymptotically anti-de Sitter 
spacetimes whose classical matter distribution is sufficiently 
well-behaved to prevent the occurence of singularities in the 
sense of null geodesic incompleteness. I'll also comment on the 
issues involved in the reconstruction of an additive and locally 
covariant bulk net of observables from a corresponding boundary net 
in this more general situation. 
\end{abstract}

\section{Introduction}

The inception of quantum field theory in curved spacetime, about
forty years ago, brought into evidence a host of new conceptual
problems hitherto absent or left unnoticed due to the peculiarities of
Minkowski spacetime, such as the very definitions of the notions
of vacuum, particle, S-matrix, etc.. One hopes that the clarification
of such issues may bring some new insights into the deeper problem of
the quantization of gravity. An example of a possible interface
between QFT in curved spacetime -- based on the \emph{principle of
locality} -- and a would-be quantum theory of gravity -- where such
principle is likely to be only macroscopically valid -- is
black hole thermodynamics. The idea that a stationary black hole is a ``black''
object in the \emph{quantum} sense of the word -- i.e., it produces
a thermal bath with a certain universal temperature -- suggests,
together with the peculiar geometrical behaviour of its event horizon,
some remarkable consequences, such as: 1.) the notion of
(thermodynamical) entropy is no longer extensive as in usual
thermodynamics, but leads to a quantity depending linearly on the
\emph{area} of the event horizon (Bekenstein-Hawking entropy)
2.) A black hole can ``evaporate'', i.e., lose all its mass by thermally
radiating it to infinity (Hawking radiation), in finite time, which leads to a
complete decoherence of an initially pure global state through entanglement
with the partial state inside the horizon of the vanishing black hole.
To explain these phenomena without violating basic postulates of
quantum mechanics, 't Hooft and Susskind have put forward the
\emph{holographic principle} -- namely, that the horizon has already
\emph{all} physical degrees of freedom, in the sense that one can
completely reconstruct the physical data contained in a (\emph{bulk}) volume from
the physical system living on the \emph{boundary} of this volume, in 
pretty much the same way a tridimensional picture is rebuilt 
from a two-dimensional hologram.\\

A concrete implementation of this principle in string theory was
conjectured by Maldacena\cite{malda} and Witten\cite{witt} -- the
famous \emph{AdS-CFT correspondence}, which triggered an impressive
amount of theoretical development afterwards. Surprisingly,
although it was thought the holographic principle to be inconsistent
with the principle of locality\cite{bousso}, it is possible to 
\emph{rigorously} prove that the essentials of the AdS-CFT 
correspondence -- more precisely, the peculiar geometry of the 
spacetimes involved -- allow the reformulation of this 
correspondence in a manner \emph{consistent with the principle 
of locality}, that is, within the context of QFT in curved 
spacetime. Such a result is proven in Rehren's paper\cite{rehren1}, 
which is formulated within the framework of Local Quantum Physics
(Algebraic Quantum Field Theory). It states that ``theories of local observables
in Anti-de Sitter ($AdS$) spacetime that are covariant under global (rigid) 
isometries can be put in an one-to-one correspondence to theories of local
observables in $AdS$'s boundary -- that is conformal to Minkowski spacetime 
of one dimension less -- that are covariant under global (rigid) conformal
transformations''. This result did not call the attention it deserved outside
the realm of Local Quantum Physics, being misinterpreted as a ``fake proof'' of 
Maldacena's AdS-CFT correspondence, and therefore deemed useless by string 
theorists; this comes as a consequence of the fact that only \emph{rigid} 
isometries are implemented, i.e., the quantum observables are completely 
decoupled from the gravitational degrees of freedom -- there's no clue to 
how the bulk quantum system transforms under arbitrary diffeomorphisms 
of spacetime, let alone how it reacts to
arbitrary, but compactly localized, changes of the metric, and how these 
changes manifest themselves in the holographic dual theory. \\

Here we understand Rehren's theorem -- called heretofore \emph{algebraic 
holography} or \emph{Rehren duality} -- as an \emph{independent} result,
that, at the same time, poses questions with a counterpart in the ``stringy'' 
AdS-CFT correspondence, and issues deeply related to the foundations of
relativistic quantum theory itself. It's from this perspective that the
author's work starts. \\

In Section \ref{sec2}, after recalling some basic definitions and 
results in Lorentzian geometry that will be needed in the sequel, we'll
extend Rehren's geometrical setup to asymptotically simple, asymptotically
AdS spacetimes of any dimension greater than 2, based in the simple, but
crucial remark: wedges in AdS are simply diamonds with both tips belonging
to the conformal infinity. This not only renders Rehren duality quite 
natural, but also shows that it depends \emph{essentially} on the global
causal structure of AdS's conformal infinity, therefore begging for a 
generalization to spacetimes who share these properties. We'll see, however,
in Section \ref{sec3}, that there are some subtle, but important aspects
in this more general setting. Namely, one need some global constraints on the
classical matter distribution (which can be put into strictly geometrical terms)
in order to algebraic holography to preserve causality when going from the 
bulk to the boundary. We'll see that these conditions also open up the
possibility of encoding bulk gravitational effects in a non-geometrical 
way at the boundary -- namely, in the form of spontaneous symmetry
breaking (breakdown of Haag duality for diamonds at the boundary), 
if the bulk theory is causal and Haag dual. This has remarkable
consequences, due to previous results by Brunetti, Guido and Longo 
about modular covariance in conformal QFT\cite{longo}.
These same conditions raise, on the other hand, great 
difficulties when it comes to reconstruct the (compact) localization of 
the bulk observables using only boundary CFT data and the $(\mbox{bulk wedge}
\Leftrightarrow\mbox{boundary diamond})$ correspondence. It can be shown, 
nevertheless, that for sufficiently small bulk diamonds this reconstruction can
indeed be done. This is just enough for \emph{additive} bulk theories,
which can thus be ``holographically rebuilt''. Section \ref{sec4} closes
with some remarks on open problems and further work to be done by the author.\\

The developments to be presented in what follows are, first and foremost, geometric.
We'll center in two essential aspects: causality and localization (in the sense
of manifold topology -- see Section \ref{sec3}). We'll refrain
from studying covariance aspects of our construction in detail, as they demand 
a separate paper of their own for a proper discussion -- we'll limit
ourselves to some remarks at the end.\\

\section{\label{sec2} Doing away with coordinates in Rehren duality}

\subsection{Some tools in Lorentzian geometry}

Let's recapitulate some definitions. For details, see the monographs of 
Wald\cite{wald}, Hawking and Ellis\cite{hawkellis},
O'Neill\cite{oneill} and Beem, Ehrlich and Easley\cite{gr}.
By a \emph{spacetime} it will
be understood a pair $(\widehat{\mathscr{M}},\widehat{g})$\footnote{The use of
hats follows the convention for the interior of the conformal completion (see later).}, 
where $\widehat{\mathscr{M}}$ is a paracompact, connected and orientable $\mathscr{C}^\infty$ 
manifold, and $\widehat{g}$ is a time-orientable, Lorentzian $\mathscr{C}^\infty$ 
metric, with Levi-Civita connection $\nabla_a$.\\

Let $\mathscr{U}\subset\widehat{\mathscr{M}}$ be an open set, and $p\in\mathscr{U}$. 
The \emph{chronological} (resp. \emph{causal}) \emph{future} of $p$ with respect to 
$\mathscr{U}$, denoted by $I^+(p,\mathscr{U})$ (resp. $I^+(p,\mathscr{U})$) is given by the 
following sets:

\begin{eqnarray}
I^+(p,\mathscr{U}) & \dot{=} & \{x\in\mathscr{U}:\exists\gamma:[0,a]\stackrel{
\mathscr{C}^\infty}{\longrightarrow}\mathscr{U}\;\mbox{timelike}\nonumber\\ & & 
\mbox{and future such that}\;\gamma(0)=p,\,\gamma(a)=x\};\label{ads1}\\
J^+(p,\mathscr{U}) & \dot{=} & \{x\in\mathscr{U}:x=p\;\mbox{ou}\;\exists\gamma:
[0,a]\stackrel{\mathscr{C}^\infty}{\longrightarrow}\mathscr{U}\;\mbox{causal}\nonumber\\
& & \mbox{and future such that}\;\gamma(0)=p,\,\gamma(a)=x\}\label{ads2}.
\end{eqnarray}

Exchanging future with past, one can define in a dual fashion the \emph{
chronological} (resp. \emph{causal}) \emph{past} $I^-(p,\mathscr{U})$ (resp. $J^-(p,
\mathscr{U})$) of $p$ with respect to $\mathscr{U}$. It follows from these definitions
that $I^\pm(p,\mathscr{U})$ is open and $int(J^\pm(p,\mathscr{U}))=I^\pm(p,
\mathscr{U})$. Using such sets we can define chronology and
causality relations between two points. Let $p,q\in\mathscr{U}\subset\mathscr{V}$.
We say that $p$ \emph{chronologically} (resp. \emph{causally}) \emph{precedes} 
$q$ with respect to $\mathscr{U}$ if $p\in I^-(q,\mathscr{U})$ (resp.
$p\in J^-(q,\mathscr{U})$). We denote this relation by $p\ll_\mathscr{U}q$
(resp. $p\leq_\mathscr{U}q$). If $p\leq_\mathscr{U}q$ and $p\neq q$, we write
$p<_\mathscr{U}q$. If
$p\ll_\mathscr{U}$\hspace*{-0.6cm}$/$\hspace*{0.5cm}$q$
and $p\gg_\mathscr{U}$\hspace*{-0.6cm}$/$\hspace*{0.5cm}$q$ 
(resp. $p\nless_\mathscr{U}q$ e $p\ngtr_\mathscr{U}q$),
we say that $p$ and $q$ are \emph{chronologically} (resp. \emph{causally}) 
\emph{disjoint} -- in this case, we write $p\curlywedge_\mathscr{U} q$
(resp. $p\perp_\mathscr{U}q$). All chronology and causality relations defined above, 
as well as the notions of chronological/causal future/past, are defined for
arbitrary nonvoid sets in an obvious way. 
If $p,q\in\mathscr{U}\subset\mathscr{V}$ implies that $p\curlywedge_\mathscr{U} q$
(resp. $p\perp_\mathscr{U}q$), we say that $\mathscr{U}$ is \emph{achronal}
(resp. \emph{acausal}) with respect to $\mathscr{U}$. For instance, given
$\mathscr{O}\subset\mathscr{U}$, the set $\partial I^{+/-}(\mathscr{O},\mathscr{U})=\partial J^{+/-}
(\mathscr{O},\mathscr{U})$ is said to be the \emph{future/past achronal boundary}
of $\mathscr{O}$ with respect to $\mathscr{U}$, and constitutes an achronal, topological 
submanifold of $\mathscr{V}$ such that every $p\in\partial I^{+/-}(\mathscr{O},\mathscr{U})$ 
belongs to a (necessarily unique) null geodesic segment, achronal with
respect to $\mathscr{U}$ and contained in $\partial I^{+/-}(\mathscr{O},\mathscr{U})$, which is 
either past/future inextendible or possesses a past/future endpoint in $\mathscr{O}$. 
Such geodesics are the \emph{null generators} of $\partial I^{+/-}(\mathscr{O},
\mathscr{U})$.\\

Let now $\mathscr{S}\subset\mathscr{V}$ be closed and achronal. The \emph{future} 
(resp. \emph{past}) \emph{domain of dependence} of $\mathscr{S}$, denoted by
$D^+(\mathscr{S})$ (resp. $D^-(\mathscr{S})$) is given by:

\begin{eqnarray}
D^{+/-}(\mathscr{S})\dot{=}\{p\in\mathscr{V}:\forall\gamma:[0,a)\longrightarrow\mathscr{V}
\mbox{ past/future inextendible,}\nonumber\\\mbox{causal such that
  }\gamma(0)=p,\,\exists b<a\mbox{ such that }\gamma(b)\in\mathscr{S}\}.\label{ads8}
\end{eqnarray}

$D(\mathscr{S})\doteq D^+(\mathscr{S})\cup D^-(
\mathscr{S})$ is said to be the \emph{domain of
dependence} or \emph{Cauchy development} of 
$\mathscr{S}$. The \emph{edge} of $\mathscr{S}$ (notation: 
$\dot{\mathscr{S}}$) is given by the points $p\in\mathscr{S}$ such that
\emph{any} open neighborhood of $p$ possesses points 
$q\in I^-(p),\,r\in I^+(p)$ and a timelike curve $\gamma$
linking $q$ to $r$, and with empty intersection with $\mathscr{S}$. 
If $S$ is not only achronal but also acausal, 
then the set $int(D(\mathscr{S}))$ 
is globally hyperbolic (in such a case, we say that
$S$ is a \emph{Cauchy surface} for $int(D(\mathscr{S}))$). 
The closed, achronal set $H^+(\mathscr{S})\doteq \overline{D^+(\mathscr{S})}\setminus
I^-(D^+(\mathscr{S}))$, denoted \emph{future Cauchy horizon} of $\mathscr{S}$, possess
the following property: any $p\in H^+(\mathscr{S})$ is contained in an achronal, 
null geodesic segment contained in $H^+(\mathscr{S})$, which is either past inextendible
or has a past endpoint in $\dot{\mathscr{S}}$. An analogous property holds for
$H^-(\mathscr{S})$, the \emph{past Cauchy horizon} of $\mathscr{S}$; the 
\emph{(full) Cauchy horizon} $H(\mathscr{S})\doteq H^+(\mathscr{S})\cup H^-(\mathscr{S})$
equals $\partial D(\mathscr{S})$.\\

Now, for the notion of \emph{conformal infinity}:

\begin{definition}
\label{inf1} {\upshape\quad The \emph{conformal infinity} or 
\emph{conformal boundary} of a $n$-di\-men\-sio\-nal spacetime $(\widehat{\mathscr{M}},
\widehat{g})$ is a $n-1$-dimensional semi-Riemannian manifold $(\mathscr{I},b)$
such that there exists a $n$-dimensional, Lorentzian manifold-with-boundary 
$(\mathscr{M},g)$ (the \emph{conformal closure} or \emph{conformal completion} of
$(\widehat{\mathscr{M}},\widehat{g})$) satisfying:
\begin{itemize}
\item $\mathscr{I}\equiv\partial\mathscr{M}$; there is a diffeomorphism $\Phi$
of $\widehat{\mathscr{M}}$ onto $\Phi(\widehat{\mathscr{M}})=\mathscr{M}
\setminus\partial\mathscr{M}$;
\item $b$ is the (possibly degenerate) semi-Riemannian metric induced
by $g$ in $\mathscr{I}$;
\item There exists a \emph{conformal (Weyl) factor} (that is, a 
real-valued, positive $\mathscr{C}^\infty$ function $\Omega$ in $\widehat{\mathscr{M}}$, 
that admits a $\mathscr{C}^\infty$ extension to $\mathscr{M}$ such that $\Omega
\!\upharpoonright_\mathscr{I}\equiv 0$ and $d\Omega\!\upharpoonright_\mathscr{I}\neq 0$ 
em $\mathscr{I}$) satisfying 
$g_{ab}=\Omega^2\widehat{g}_{ab}$. 
\end{itemize}}
\end{definition}

Note that, if $n_a:=\nabla_a\Omega\neq 0$ in $\mathscr{I}$ ($n_a$ is the \emph{normal 
(co)vector} to $\mathscr{I}$; this condition can be made to hold if the Einstein equations
are satisfied in a neighborhood of $(\mathscr{I},b)$ and the
classical matter fields satisfy certain decay conditions in this neighborhood),
then $\Omega$ can be chosen in such a way that the extrinsic curvature (second
fundamental form) $K_{ab}:=\nabla_a n_b$ vanishes in $\mathscr{I}$. It is enough to multiply
$\Omega$ by a real-valued, positive, nowhere vanishing $\mathscr{C}^\infty$ 
function $\omega$ in $\mathscr{M}$ -- the new factor $\Omega$ still satisfies all the conditions in
Definition \ref{inf1}. Nonetheless, this choice by no means constrains the values
$\omega$ can take in $\mathscr{I}$\cite{ashtekar}. Therefore, $(\mathscr{I},b)$ 
\emph{can be taken totally geodesic, regardless of the choice of representative 
of the conformal structure of $b$}.\\

\begin{definition}
\label{inf2} {\upshape\quad Let $(\widehat{\mathscr{M}},\widehat{g})$  
be a $n$-dimensional spacetime with conformal infinity $(\mathscr{I},
b)$. We say that $(\widehat{\mathscr{M}},\widehat{g})$ is \emph{
asymptotically simple} if any inextendible null geodesic $(\widehat{\mathscr{M}},\widehat{g})$
has an unique extension to $(\mathscr{M},g)$ such that
 $\mathscr{I}$ contains precisely its both endpoints.}
\end{definition}

Obviously, this is only possible if $(\widehat{\mathscr{M}},\widehat{g})$
is \emph{null geodesically complete}. Actually, when $(\mathscr{I},
b)$ is timelike (and therefore a spacetime in its own right), one can say more, 
justifying the name ``asymptotically simple'':

\begin{theorem}
\label{t1} {\upshape\quad If $(\widehat{\mathscr{M}},\widehat{g})$ is
asymptotically simple and has a timelike conformal infinity, then it 
is causally simple, that is, $J^\pm(p,\widehat{\mathscr{M}})$ is
closed in $\widehat{\mathscr{M}}$ (and therefore equal to $\overline{
I^\pm(p,\widehat{\mathscr{M}})}$) for all $p\in\widehat{\mathscr{M}}$.}
\begin{quote}{\small\scshape Proof.\quad}
{\small\upshape First, notice that if $p,q\in\widehat{\mathscr{M}}$ are such that
$p\ll_{\widehat{\mathscr{M}}}$\hspace*{-0.6cm}$/$\hspace*{0.5cm}$q$.
then $p\ll_{\mathscr{M}}$\hspace*{-0.6cm}$/$\hspace*{0.5cm}$q$ and likewise 
exchanging future with past, for a timelike curve in $\mathscr{M}$ linking 
$p$ to $q$ can always be slighly deformed so as to give a timelike curve
\emph{contained in $\widehat{\mathscr{M}}$} and linking $p$ to $q$.
Now, suppose that $p\in\partial I^-(q,\widehat{\mathscr{M}})$ and
$p\notin J^-(q,\widehat{\mathscr{M}})$. By the reasoning above,
we have $p\in\partial I^-(q,\mathscr{M})$. Moreover, by hypothesis, 
a null generator $\gamma$ of $\partial I^-(q,\widehat{\mathscr{M}})$ must
reach its future endpoint at infinity without crossing $q$ before this. Let $r$ be 
such an endpoint. Then, $r\in\partial I^-(q,\mathscr{M})$ since this set is closed. 
Since the infinity is totally geodesic, $\gamma$ must hit it transversally and
thus any future causal extension of $\gamma$ must be broken. Therefore, if one
extends $\gamma$ slightly to the future by a null generator segment $\gamma'$ of $\partial 
I^-(q,\mathscr{M})$ crossing $r$ (say, by setting the affine parameter $t$ of $\gamma'$ equal
to zero in $r$ and extending up to $t=\epsilon>0$), then there is a timelike 
curve in $\mathscr{M}$ linking $p$ to $\gamma'(\epsilon)$\cite{hawkellis}, which
violates the achronality of $\partial I^-(q,\mathscr{M})$. Repeat the argument exchanging
future with past.\hfill$\Box$}\end{quote}
\end{theorem}

An asymptotically simple spacetime, however, need not be globally hyperbolic --
a prime example is AdS spacetime, which will be studied in the next Subsection.

\subsection{Asymptotically Anti-de Sitter (AAdS) spacetimes and Rehren duality}

We'll recapitulate some definitions given in \cite{ribeiro2}.
Recall that $n$-dimensional AdS spacetime (notation: $AdS_n$, $n\geq 3$) is 
given by the hyperquadric in $\mathbb{R}^{n+1}$ ($\mathbf{X}\,=\,
(X^1,X^2,\ldots,X^{n-2})$)
\begin{equation}\label{ads10}
-X^0X^0+\mathbf{X\cdot
  X}+X^{n-1}X^{n-1}-X^nX^n=A^2,\,A>0,
\end{equation}
where the $X^0-X^n$ plane determines the time orientation. $AdS_n$ is
an homogeneous space for the isometry group $SO(2,n-1)$. 
Consider now the region $AdS^+_n\doteq\{X\in AdS_n:X^{n-1}+X^n>0\}$. One can 
build a chart for this region (denoted \emph{horocyclic} or \emph{Poincaré
parametrization}) with the parameters $(x,z)$, where $x\in\mathbb{R}^{1,n-2}$ 
and $z\in\mathbb{R}_+$:
\begin{equation}\label{ads11}
\left\{ \begin{array}{l@{=}l}
X^\mu & \frac{A}{z}x^\mu\;(\mu=0,\ldots,n-2) \\
\frac{1}{A}X^{n-1} & \frac{1-z^2}{2z}+\frac{1}{2z}x_\mu x^\mu \\
\frac{1}{A}X^n & \frac{1+z^2}{2z}-\frac{1}{2z}x_\mu x^\mu
\end{array} \right.
\end{equation}

One can see from the formulae (\ref{ads11}) that each timelike hypersurface
given by $z=const.$ is conformal to $\mathbb{R}^{1,n-2}$ by a factor
\begin{equation}\label{ads12}
(X^{n-1}+X^n)^2=\frac{A^2}{z^2}. 
\end{equation}

In this chart, the $AdS_n$ metric is written as

\begin{equation}\label{ads13}
ds^2\,=\,\frac{A^2}{z^2}(dx_\mu dx^\mu+dz^2),
\end{equation}
that is, $AdS^+_n$ is a semi-Riemannian \emph{``warped product''}
of $\mathbb{R}^{1,n-2}$ with $\mathbb{R}^*_+$. The universal covering
of $AdS_n$ (notation: $\widetilde{AdS_n}$) is asymptotically simple, and possesses 
the Einstein static universe (ESU) $\mathscr{I}=\mathbb{R}\times S^{n-2}$ as conformal infinity. 
Specializing to the Poincaré chart, we see that in $AdS^+_n$ the conformal factor is
given by (\ref{ads12}), that is, $z=0$ corresponds to the conformal embedding
of Minkowski spacetime into the Einstein static universe. $\widetilde{AdS_n}$
satisfies the empty space Einstein equations with (negative) cosmological 
constant $\Lambda=-\frac{(n-1)(n-2)}{2A^2}$.\\

Now, let us define, for $p,q\in\mathscr{I},p\ll_\mathscr{I} q$:

\begin{eqnarray}
\mathscr{W}_{p,q} & \doteq & (I^-(p,\overline{AdS_n})\cap I^+(q,\overline{
AdS_n}))\cap\widetilde{AdS_n}\;\mbox{ ((bulk) \emph{wedge})};\label{ads14}\\
\mathscr{D}_{p,q} & \doteq & (I^-(p,\overline{AdS_n})\cap I^+(q,\overline{
AdS_n}))\cap\mathscr{I}=\nonumber\\ & = & I^-(p,\mathscr{I})\cap I^+(q,\mathscr{I})
\mbox{ ((boundary) \emph{diamond})}.\label{ads15}
\end{eqnarray}

Let $p\in\mathscr{I}$. All future null geodesics emanating from
$p$ will focus at a single point of $\mathscr{I}$, which is the future
endpoint of all null generators of $\partial I^+(p,\mathscr{I})$. This point is denoted 
\emph{antipodal} of $p$ (notation: $\bar{p}$). The antipodal of $p$ has the following properties:

\begin{eqnarray}
\partial I^+(p,\mathscr{I})=\partial I^-(\bar{p},\mathscr{I});\label{ads16}\\
\partial I^+(p,\overline{AdS_n})=\partial I^-(\bar{p},\overline{AdS_n}).\label{ads17}
\end{eqnarray}

Let $\bar{\bar{p}}\doteq\overline{(\bar{p})}$, and define
$\mathscr{M}in(p)\doteq\mathscr{D}_{p,\bar{\bar{p}}}$, 
the \emph{Minkowski domain} to the future of $p\in\mathscr{I}$.
This region corresponds to the conformal embedding of $\mathbb{R}^{1,n-2}$
into $\mathscr{I}$ such that $p$ corresponds to the past timelike infinity of $\mathbb{R}^{1,n-2}$. 
$\mathscr{P}oi(p)\doteq\mathscr{W}_{p,\bar{\bar{p}}}$ 
corresponds to the domain of a Poincaré chart in $\widetilde{AdS_n}$, 
therefore being denominated \emph{Poincaré domain} to the future of $p$. Given 
the objects defined above, we can define the geometrical setup for Rehren duality
as follows:

\begin{enumerate}
\item The isometry group of $\widetilde{AdS_n}$ \emph{acts transitively} on the collections
$\mathfrak{W}\doteq\{\mathscr{W}_{p,q}:p,q\in\mathscr{I}\}$ of bulk wedges and
$\mathfrak{D}\doteq\{\mathscr{D}_{p,q}:p,q\in\mathscr{I}\}$
of boundary diamonds.
\item $\mathscr{W}_{p,q}$ and $\mathscr{D}_{p,q}$ share the \emph{same isotropy subgroup}.
\item From (\ref{ads16}) and (\ref{ads17}), it follows respectively that 
$\mathfrak{W}$ and $\mathfrak{D}$ are closed under causal complements. More precisely
(by $\mathscr{O}'_{\mathscr{U}}$ we mean the causal complement of $\mathscr{O}$ 
with respect to $\mathscr{U}\supset\mathscr{O}$), we have, for all 
$ p,q\in\mathscr{M}in(r),\,r\in\mathscr{I},\,p\ll_\mathscr{I}\bar{q}$, 
\begin{equation}\label{ads18}
\overline{\mathscr{W}_{q,\bar{p}}}\cap\widetilde{AdS_n}=(\mathscr{W}_{p,\bar{q}})'_{\widetilde{AdS_n}}
\end{equation}
and
\begin{equation}\label{ads19}
\overline{\mathscr{D}_{q,\bar{p}}}=(\mathscr{D}_{p,\bar{q}})'_\mathscr{I}.
\end{equation}
\item As a consequence of the above statements, the \emph{Rehren bijection} 
\begin{eqnarray}
\rho:\mathfrak{W} & \longrightarrow & \mathfrak{D}\label{ads20}\\
\mathscr{W}_{p,q} & \mapsto & \alpha(\mathscr{W}_{p,q})\doteq\mathscr{D}_{p,q}\nonumber
\end{eqnarray}
is one-to-one and onto, preserves inclusions and causal complements, and
intertwines the action of $\widetilde{AdS_n}$'s isometry group, which is also
the conformal group of $(\mathscr{I},b)$ and the universal covering of the conformal
group of Minkowski spacetime. \emph{Rehren duality} = \emph{algebraic holography} is 
simply the \emph{transplantation} (change of index set) of theories of local 
observables under the map $\rho$\footnote{This coordinate-free form of the Rehren bijection, 
which solely makes use of causal relationships in the conformal closure, is based 
in a suggestion from prof. K.-H. Rehren\cite{rehren2}, and was employed in this 
form by Bousso and Randall\cite{boura} for studying qualitative aspects of the 
AdS-CFT correspondence.}.
\end{enumerate}

Now, what happens if we ``perturb'' AdS spacetime in such a way that \emph{we
still have the ESU $(\mathscr{I},b)$ as conformal infinity?} This correspond to 
the class of \emph{asymptotically AdS} spacetimes. More precisely,
by employing a definition similar to the ones given in \cite{ashtekar} and \cite{ashdas},
one can write:

\begin{definition}
\label{inf3} {\upshape\quad A $n$-dimensional spacetime ($n\geq 3$)
$(\widehat{\mathscr{M}},\widehat{g})$ with conformal infinity $(\mathscr{I},b)$
is said to be \emph{asymptotically anti-de Sitter} (notation: AAdS) if:
\begin{enumerate}
\item It satisfies Einstein's equations $\widehat{R}_{ab}-\frac{1}{2}
\widehat{R}\widehat{g}_{ab}-\Lambda\widehat{g}_{ab}\,=\,8\pi G_{(n)}\widehat{T}_{ab}$,
where $G_{(n)}$ is the $n$-dimensional Newton's constant, and the cosmological constant 
$\Lambda$ is $<0$ (one can attribute an ``AdS radius'' to such spacetimes, by setting
$A=\sqrt{-\frac{(n-1)(n-2)}{\Lambda}}$);
\item $(\mathscr{I},b)$ is globally conformally diffeomorphic to the $(n-1)$-dimensional
Einstein static universe;
\item The (classical) energy-momentum tensor $\widehat{T}_{ab}$ of 
$(\widehat{\mathscr{M}},\widehat{g})$ decays fast enough close to
$\mathscr{I}$ for $\Omega^{2-n}\widehat{T}^a_b$ to possess a $\mathscr{C}^\infty$
extension to the conformal closure $(\mathscr{M},g)$.
\end{enumerate}}
\end{definition}

The condition on the decay of $\widehat{T}_{ab}$ is motivated by considering
the behaviour of classical fields emanating from compactly localized sources in 
$AdS_n$, specially massless fields (electromagnetic, Yang-Mills). The global condition
on the conformal infinity makes sense in general because solutions of Einstein's 
equations with negative cosmological constant possess a timelike conformal infinity.\\

In what follows, we shall make two additional demands on the class of AAdS spacetimes we'll
deal with:

\begin{description}
\item[\textbf{Asymptotic simplicity.}] This is indispensable for rebuilding bulk localization
from wedges. The existence of a large class of asymptotically simple AAdS spacetimes was
proven by Friedrich\cite{friedrich1,friedrich3}.\\
\item[\textbf{Global focusing of null geodesics.}] More precisely, it's demanded that all
inextendible null geodesics shall possess a pair of conjugate points
(recall that a pair of points $p,q$ in a null geodesic $\gamma$ are said to be
\emph{conjugate} if there's a Jacobi field on $\gamma$ -- i.e., a vector field that
satisfies the geodesic deviation equation on each point of $\gamma$ -- nowhere
vanishing on the open segment of $\gamma$ linking $p$ to $q$ but vanishing at
both $p$ and $q$. It's well known\cite{gr,wald,hawkellis} that, in this case,
any point of $\gamma$ to the future (resp. past) of $q$ (resp. $p$) can be 
linked to $p$ (resp.$q$) by a timelike curve). It's precisely
this condition that guarantees the Rehren bijection will preserve causality, and
it also ends up playing an important role in the reconstruction of bulk localizatiion. 
Even if one does not require asymptotic simplicity, one 
can still show that any chronological spacetime
which satisfies this focusing condition is strongly causal\cite{gr}, and therefore
its topology is generated by diamonds. AdS does not satisfy this condition, but it follows
from energy conditions on the energy-momentum tensor as weak as the NEC, ANEC and the
Borde energy condition\cite{borde}\footnote{If one wants to extend our considerations to 
\emph{semiclassical} AAdS spacetimes, i.e., with quantum backreaction, we remark 
that quantum energy inequalities seem to be capable of guaranteeing that a ``Planck-scale 
coarse-grained (i.e., transversally smeared)'' ANEC holds for the renormalized quantum 
energy-momentum tensor\cite{roman}, but whether this implies, say, the Borde energy 
condition, and thus gives rise to the needed focusing theorems\cite{borde}, or not, it's 
an open question so far.}, and does look natural from the viewpoint of certain 
\emph{rigidity theorems} for asymptotically simple spacetimes: for asymptotically flat 
and de Sitter spacetimes which satisfy, say, NEC, it was proven by Galloway
\cite{galloway1,galloway2}, by employing the stability results of Friedrich
\cite{friedrich1,friedrich2}, that if such spacetimes possess a so-called 
\emph{null line} (a complete, achronal null geodesic), then they are \emph{globally 
isometric} to Minkowski spacetime (resp. de Sitter spacetime, with radius determined 
by the value of the cosmological constant appearing in the Einstein equations). From the
viewpoint of \emph{stability of the (conformal)\cite{friedrich1} mixed 
Cauchy/boundary problem},
one can see that the occurrence of null lines is an \emph{unstable} feature of 
such spacetimes, i.e., any arbitrarily small perturbation of Cauchy data that 
preserves boundary conditions at the conformal infinity \emph{destroys all null 
lines}, i.e., all complete null geodesics acquire a pair os conjugate points. There 
is still no similar result for asymptotically simple AAdS spacetimes in the sense 
of Definition \ref{inf3}, yet the the global structure of conformal infinity 
suggests that this may still be true. If so, our analysis complements and 
extends Rehren's.
\end{description}

We shall now study how this framework behaves in the more general
situation of AAdS spacetimes complying with the conditions above.

\section{\label{sec3} Properties of the Rehren bijection in AAdS spacetimes}

\subsection{\label{causal} Causality (bulk-to-boundary)}

In principle, gravitational effects deep inside the bulk may produce \emph{causal
shortcuts} through the bulk linking causally disjoint point at the boundary, i.e.,
it may happen that $I^+(p,\mathscr{I})\cap I^-(q,\mathscr{I})\subsetneqq 
(I^+(p,\mathscr{M})\cap I^-(q,\mathscr{M}))\cap\mathscr{I}$, rendering the second
identity in (\ref{ads15}) \emph{false}. Such a thing would be ruinous to the
Rehren bijection to keep preserving causality in AAdS spacetimes. We shall
show now that, luckily, (\ref{ads15}) still holds under our set of hypotheses.
Here, we'll make use of the notion of \emph{gravitational time delay}\cite{gao,sorkin,woolgar}
of complete null geodesics in AAdS spacetimes.
The Einstein static universe $(\mathscr{I},b)$ is globally
hyperbolic; let it be given a foliation of it in Cauchy surfaces such that the
orbits of the global time function $t$ (supposed to be oriented in the same way
as the time orientation of $(\mathscr{I},b)$) generating the foliation are complete 
timelike geodesics, and the values of the global time function correspond
to a common affine parametrization of this family of geodesics. Now, let $\gamma$
be a complete null geodesic in $\mathscr{M}$ 
traversing $\widehat{\mathscr{M}}$, with past endpoint $p$ and future endpoint $q$ 
belonging to the orbits $T_p$ (resp. $T_q$), and $\gamma'$ a null geodesic segment \emph{in}
$\mathscr{I}$ starting at $p$ and ending at, say, $q'\in T_q$. The
gravitational time delay of $\gamma$ with respect to $\mathscr{I}$ is 
given by the difference $\Delta t=t(q)-t(q')$ 
(notice that, due to the properties of null geodesics in ESU, it follows that
any other null geodesic segment in $\mathscr{I}$ starting at $p$ that crosses
$T_q$ afterwards will necessarily do it at $q'$.). Although this value depends on
the choice of foliation, the \emph{sign} of $\Delta t$ ($<0$, $=0$, $>0$) does not.
Under our set of hypotheses, the gravitational time delay in AAdS spacetimes is
always \emph{postitive}:\\

\begin{theorem}
\label{t2}\footnote{The method of proof was communicated to me by
Sumati Surya\cite{surya}. It's analogous to the proof
of a positive mass theorem for asymptotically flat spacetimes due to
Penrose, Sorkin and Woolgar\cite{sorkin} and for AAdS spacetimes due to Woolgar\cite{woolgar}.
Another proof of this, using a somewhat different strategy, was provided by
Gao and Wald\cite{gao}.} 
{\upshape\quad Let $(\widehat{\mathscr{M}},\widehat{g})$
be an asymptotically simple AAdS spacetime, such that every inextendible
null geodesic has a pair of conjugate points, and $p\in\mathscr{I}$. 
Then, evert null geodesic segment emanating from $p$ which doesn't belong to 
$\mathscr{I}$ has its future endpoint in $I^+(p,\mathscr{I})$.}
\begin{quote}{\small\scshape Proof.\quad}
{\small\upshape Let $\gamma$ be a null geodesic segment emanating from $p$ and traversing the bulk, and
let $p'$ be the future endpoint of $\gamma$ in $\mathscr{I}$. 
Since we've assumed that $(\widehat{\mathscr{M}},\widehat{g})$  
is causal, one can see that $p'\notin\overline{I^-(p,\mathscr{M})}$. 
Now, we prove two Lemmata:

\begin{lemma}
\label{l1} {\scshape (Absence of causal shortcuts) \upshape\quad  Let $p,p'\in\mathscr{I}$.  
If $p\perp_\mathscr{I}p'$, then there is no causal curve in $(\mathscr{M},g)$
linking $p$ to $p'$.}
\begin{quote}{\small\scshape Proof.\quad}
{\small\upshape Suppose that $p'>_\mathscr{M}p$ (the opposite case is treated analogously). 
We'll prove that the gravitational time delay implied by the presence of a pair of
conjugate points contradicts the causal disjointness of $p$ and $p'$ with respect to 
$\mathscr{I}$. Denote by $T(p')$ the timelike generator of $(\mathscr{I},b)$ containing
$p'$. \\

Note that $\partial I^+(p,\mathscr{I})\doteq\Sigma$ is a closed, achronal surface 
that cuts $(\mathscr{I},b)$, in two disjoint subsets
$I^+(p,\mathscr{I})\doteq A$ and $\mathscr{I}\setminus\overline{I^+(p,\mathscr{I})}\doteq B$
and intersects each timelike generator of
$\mathscr{I}$ in precisely one point, as every timelike generator has points in
$I^+(p,\mathscr{I})$ and $I^-(\bar{p},\mathscr{I})$.
By hypothesis, $p'\in B$. Moreover, $T(p')$ must cross $\Sigma$ at some instant of time. 
Therefore, there exists $p''\in T(p')$ such that $p''\gg_\mathscr{I} p'$ and 
$p''\in\Sigma$. Let $\gamma$ be a null generator of $\Sigma$ that contains $p''$. As the segment 
of $\gamma$ that links $p$ to $p''$ is null and achronal, $\gamma$ is necessarily the fastest 
curve in $(\mathscr{I},b)$ linking $p$ to $T(p')$. \\

Now, consider the achronal boundary $\partial I^+(p,\mathscr{M})=\partial J^+(p,\mathscr{M})
\doteq\overline{\Sigma}$. $\overline{\Sigma}\cap\mathscr{I}$ is
closed, achronal and intersects each timelike generator of
$\mathscr{I}$ in precisely one point, as every timelike generator has points in
$I^+(p,\mathscr{M})$ and $I^-(p,\mathscr{M})$, and $\overline{\Sigma}$ separates 
$\mathscr{M}$ in two disjoint open sets (from the viewpoint of a manifold-with-boundary,
of course) $I^+(p,\mathscr{M})\doteq\dot{A}$ and $\mathscr{M}
\setminus\overline{I^+(p,\mathscr{M})}\doteq\dot{B}$. Thus, $T(p')$ 
must cross $\overline{\Sigma}$ in, say, $p'''$. Since $p<_\mathscr{M}p'$, we must have 
$p'''\ll_\mathscr{I} p'$ or $p'''=p'$. In both cases, we have
$p'''\ll_\mathscr{I}p''$, which implies that any null generator $\overline{\gamma}$ of $\overline{\Sigma}$
containing $p'''$ is strictly faster than $\gamma$. As $\gamma$ was the fastest curve
in $(\mathscr{I},qb)$ linking $p$ to $T(p')$, $\overline{\gamma}$ necessarily
traverses $\widehat{\mathscr{M}}$. Hence, we've built a complete and achronal null geodesic
in $(\widehat{\mathscr{M}},\widehat{g})$. However, such a geodesic cannot exist
since it must have a pair of conjugate points and therefore cannot be achronal.
\hfill$\Box$}\end{quote}
\end{lemma}

\begin{lemma}
\label{l2} {\upshape\quad Let $p,p'\in\mathscr{I}$. If 
$p'\in\partial I^+(p,\mathscr{I})$ and $p'\neq \bar{p}$, then there is no null geodesic segment in 
$(\mathscr{M},g)$ that doesn't belong to $\mathscr{I}$ and links $p$ to $p'$.}
\begin{quote}{\small\scshape Proof.\quad}
{\small\upshape Let $\gamma$ be the (necessarily unique) null generator of $\partial 
I^+(p,\mathscr{I})$ linking $p$ to $p'$. Suppose that there is another null geodesic,
traversing $\widehat{\mathscr{M}}$ and linking $p$ to $p'$. Then, if one picks any point
$p''$ in $\gamma$ after $p'$, there is a \emph{broken} null geodesic segment linking
$p$ to $p''$, which in turn implies that there is a timelike curve in $(\mathscr{M},g)$
linking $p$ to $p''$\cite{hawkellis}\\

Let $T(p'')$ be the timelike generator of $(\mathscr{I},b)$ containing $p''$.
Now, consider $\overline{\Sigma}$ as in the preceding Lemma. Again, $T(p'')$ must cross
$\overline{\Sigma}$, say, in $p'''$. Thus, necessarily $p'''\ll_\mathscr{I} p''$. But this implies
that, since $(\mathscr{M},g)$ is causal, $p\perp_\mathscr{I}p'$. That contradicts
the preceding Lemma.
\hfill$\Box$}\end{quote}
\end{lemma}

Both Lemmata above imply that, if $p'$ isn't dragged inside $I^+(p,\mathscr{I})$ 
by gravitational time delay, then $p'$ coincides with $\overline{p}$, just like
in AdS. But, even in such a case, the presence of conjugate points in $\gamma$ 
implies that there is a timelike curve traversing the bulk and linking $p$ to
$\bar{p}$. Repeating the argument in Lemma \ref{l2}, the result follows.
\hfill$\Box$}\end{quote}
\end{theorem}

All arguments above can be repeated exchanging future with past, and $p$ with its
antipodal $\bar{p}$, yielding a similar result in the opposite time orientation.
It's now immediate to show that, as a direct consequence of Theorem \ref{t2},
(\ref{ads15}) holds. Thus, we shall keep the same notation for the conformal infinity, for bulk wedges,
boundary diamonds and the Rehren bijection, as well as for Minkowski and Poincaré domains, 
when dealing with AAdS spacetimes.

\begin{remark}\label{r1}
Our set of hypotheses, however, entails that, although the Rehren
bijection preserves causality, it no longer does so in a \emph{maximal} way -- the collection
of wedges is no longer closed under causal complements. More precisely:
\end{remark}

\begin{proposition}
\label{p1} {\upshape\quad Let $(\widehat{\mathscr{M}},\widehat{g})$
be an AAdS spacetime satisfying the hypotheses of Theorem \ref{t2}. Define 
$\Xi^+_p\doteq\partial I^+(p,\mathscr{M})\setminus \{p,\overline{p}\}
$ and $\Xi^-_{\overline{p}}\doteq\partial I^-(\overline{p},\mathscr{M})\setminus \{p,\overline{p}\}$. 
Then: 
\begin{enumerate}
\item[(i)] $\Xi^+_p\cap I^-(\overline{p},\mathscr{M})=\Xi^-_{\overline{p}}\cap I^+(p,\mathscr{M})=
\varnothing$.
\item[(ii)] $\Xi^+_p\cap\Xi^-_{\overline{p}}\cap\widehat{\mathscr{M}}=\varnothing$.
\end{enumerate}}
\begin{quote}{\small\scshape Proof.\quad}
{\small\upshape (i) We know that $\Xi^+_p\cap\mathscr{I}=\Xi^-_{\overline{p}}\cap\mathscr{I}$, 
so let's concentrate only at the bulk. Namely, suppose that there is $q\in\widehat{\mathscr{M}}$
such that $q\in\Xi^-_{\overline{p}}$ and $q\notin\Xi^+_p$. If $q\gg p$, that contradicts
the fact that there is no timelike curve linking $p$ to $\overline{p}$. Repeat
the argument exchanging past with future, and the roles of $p$ and $\overline{p}$. (ii) 
If $\Xi^+_p$ and $\Xi^-_{\overline{p}}$ coincide in some point $q\in\widehat{M}$, 
then there is an at least broken null geodesic segment linking $p$ to $\bar{p}$, 
which implies that there is a timelike curve through the bulk doing the same.
The result easily follows by repeating the reasoning at the end of the proof of
Theorem \ref{t2}.
\hfill$\Box$}\end{quote}
\end{proposition}

Note that the argument above is symmetric with respect to time orientation.
An important consequence of Proposition \ref{p1} is that our global focusing
condition entails a nontrivial ``shrinking'' of the  
AAdS wedges towards the boundary, as a side effect of the gravitational time delay. 
As a consequence of the latter, it follows that, although by Theorem \ref{t2} 
$\bar{p}$ still satisfies (\ref{ads16}), it will certainly 
violate (\ref{ads17}). Moreover, $\overline{\mathscr{W}_{q,\bar{p}}}
\cap\widehat{\mathscr{M}}\subsetneqq(\mathscr{W}_{p,\bar{q}})'_{\widehat{
\mathscr{M}}}$ and $\mathscr{P}oi(r)\subsetneqq\{r\}'_{\mathscr{M}}\cap
\widehat{M}$. Hence, property 3 for wedges no longer remains valid, and therefore
the collection of bulk wedges cannot be closed by causal complements in the sense of 
(\ref{ads18}). The strict inclusions above suggest that Haag duality for boundary diamonds 
cannot be satisfied \emph{without violating bulk causality} for local observables,
as the local (von Neumann, for concreteness) algebra $\mathfrak{A}(\mathscr{W}
(p,\bar{q})')$ may be \emph{strictly larger} than $\mathfrak{A}(\mathscr{W}(q,\bar{p}))=
\mathfrak{A}(\mathscr{D}(q,\bar{p}))$ (here the Borchers timelike tube theorem doesn't
remove the strictness of the inclusion, as here one would need to extend the localization
beyond the set of points causally between $p$ and $\bar{q}$). Hence, in such a case,
if $\mathfrak{A}(\mathscr{D}(p,\bar{q}))'
=\mathfrak{A}(\mathscr{D}(q,\bar{p}))=\mathfrak{A}(\mathscr{D}(p,\bar{q})')$, then
the algebra $\mathfrak{A}(\mathscr{W}(p,\bar{q})')$ necessarily has elements that
do not commute with $\mathfrak{A}(\mathscr{W}(p,\bar{q}))=\mathfrak{A}(
\mathscr{D}(p,\bar{q}))$, therefore violating local causality. 
That means that the local algebras also ``shrink'' from the viewpoint of the boundary -- 
in this way, the boundary net can ``feel'' bulk gravitational effects
as a spontaneous symmetry breaking that necessarily follows
from the breakdown of Haag duality\cite{haag} -- more on this at the end of this
contribution.

\subsection{\label{envelope} Reconstruction (boundary-to-bulk)}

Knowing the wedge localization of bulk observables may not be enough
for the full reconstruction of the bulk quantum theory using only boundary
data and the Rehren bijection $\rho$. We need to be able to specify the
localization of the local procedures in arbitrarily small open regions, or,
which amounts to the same thing, the localization with respect to a basis
of the bulk topology. This can, in principle, be performed by taking intersections
of wedges, but it's by no means clear whether these give a basis for the bulk 
topology or not. This must be done in a more precise way. In AdS, we're fortunate, because
any \emph{relatively compact AdS diamond} (for the purpose of generating the manifold topology,
these suffice) \emph{can be enveloped by AdS wedges}: given
\begin{equation}\label{ads21}
\mathscr{O}_{p,q}\doteq I^+(p,\widetilde{AdS_n})\cap I^-(q,\widetilde{AdS_n}),
\,p\ll_{\widetilde{AdS_n}}q,\end{equation}
we can write
\begin{equation}\label{ads22}
\mathscr{O}_{p,q}=\bigcap_{r\in\partial I^-(p,\overline{AdS_n})\cap\mathscr{I},\,
s\in\partial I^+(q,\overline{AdS_n}),\,r,s\in\mathscr{M}in(u)}\mathscr{W}_{r,s}.
\end{equation}

We'll see shortly that the achronality of the inextendible null geodesics
in AdS is crucial for the precise enveloping.
In an AAdS spacetime as in Theorem \ref{t2}, the issue is much more delicate, 
because of the following
 
\begin{proposition}
\label{p2} {\upshape\quad Let $(\widehat{\mathscr{M}},\widehat{g})$
be an AAdS spacetime satisfying the hypotheses of Theorem \ref{t2}, and
$p\in\widehat{\mathscr{M}}$. Then, $\partial I^+(p,\mathscr{M})$ intersect
each timelike generator if $(\mathscr{I},b)$ \emph{precisely once}.}
\begin{quote}{\small\scshape Proof.\quad}
{\small\upshape By the achronality of $\partial I^+(p,\mathscr{M})$,
it intersects every timelike generator of $(\mathscr{I},b)$ 
\emph{at most once}. Suppose that the thesis is false. Then, given a timelike
generator $T$, we have the following possibilities:
\begin{enumerate}
\item[(i)] \fbox{$T\subset I^+(p,\mathscr{M})$} -- Consider a complete null 
geodesic $\gamma$ crossing $p$, and let $q$ be the \emph{past} endpoint of 
$\gamma$. Then, there exists a value $t$ of the affine parameter of $T$ such 
that $T(t)\ll_\mathscr{I}q$. Therefore, $T(t)\ll_\mathscr{M}p$, which is absurd 
since $(\mathscr{M},g)$ is chronological.
\item[(ii)] \fbox{$T\cap J^+(p,\mathscr{M})=\varnothing$} -- Consider 
a complete null geodesic $\gamma$ crossing $p$, and let $r$ be the \emph{future}
endpoint of $\gamma$. Then, there exists a value $t$ of the affine parameter of $T$ such 
that $T(t)\gg_\mathscr{I}r$. Therefore, $T(t)\gg_\mathscr{M}p$, contradicting 
the hypothesis.
\end{enumerate}
\hfill$\Box$}\end{quote}
\end{proposition}

\begin{proposition}
\label{p3} {\upshape\quad Let $(\widehat{\mathscr{M}},\widehat{g})$ be
an AAdS spacetime which satisfies the hypotheses of Theorem \ref{t2}, 
$q,r\in\widehat{\mathscr{M}}$ such that $r\in\partial I^-(q,\widehat{\mathscr{M}})$, and
$\gamma$ a null generator of $\partial I^-(q,\widehat{\mathscr{M}})$ to which $r$ belongs. Let $s_1(r),s_2(r),
s_3(r)\in\mathscr{I}$ be defined as:
\begin{itemize}
\item $s_1(r)$ is the future endpoint of $\gamma$;
\item $s_2(r)$ is the point where $\partial I^+(q,\mathscr{M})$ intersects
the timelike generator $T(s_1(r))$ to which $s_1(r)$ belongs;
\item $s_3(r)$ is the point where $\partial I^+(r,\mathscr{M})$ intersects
$T(s_1(r))$.
\end{itemize}
Then:
\begin{enumerate}
\item[(i)] $s_3(r)\leq_\mathscr{I}s_2(r)\leq_\mathscr{I}s_1(r)$.
\item[(ii)] $s_3(r)=s_2(r)=s_1(r)$ if and only if the segment of $\gamma$ linking
$r$ to $s_1(r)$ is achronal.
\end{enumerate}}
\begin{quote}{\small\scshape Proof.\quad}
{\small\upshape (i) Immediate, as is (ii) $\Rightarrow$. It remains to prove 
(ii) $\Leftarrow$. Namely, suppose that $s_3(r)$ equals $s_2(r)$. Then, the
null geodesic segment linking $q$ to $s_2(r)$ must belong to $\gamma$, for otherwise
there would be a broken geodesic segment linking $r$ and $s_3(r)$, contradicting 
the definition of the latter (this, in particular, proves that $s_1(r)=s_3(r)$ even
if we just assume $s_2(r)=s_3(r)$). If $\gamma$ is not achronal, once more we have a 
contradiction with the definition of $s_3(r)$.\hfill$\Box$}\end{quote}
\end{proposition}

Similar results are valid if we exchange future with past.
Now, in an AAdS spacetime satisfying the hypotheses of
Theorem \ref{t2}, let $\mathscr{O}_{p,q}$ be a \emph{relatively compact diamond
with a contractible Cauchy surface} -- any sufficiently small diamond satisfies
both conditions. Let's now consider the region

\begin{equation}\label{ads23}
\mathscr{Q}_{p,q}=\bigcap_{r\in\partial I^-(p,\mathscr{M})\cap\mathscr{I},
s\in\partial I^+(q,\mathscr{M})\cap T(r)}\mathscr{W}_{r,s}
\end{equation}
It follows naturally from the definition that $\mathscr{Q}_{p,q}\supset\mathscr{O}_{p,q}$, 
it is \emph{causally complete}, as it is the intersection of causally complete regions,
and

\begin{equation}\label{ads24}
\mathscr{Q}_{p,q}\cap J^+(q,\mathscr{M})=\mathscr{Q}_{p,q}\cap 
J^-(p,\mathscr{M})=\varnothing.
\end{equation}

In AdS, $\mathscr{Q}_{p,q}=\mathscr{O}_{p,q}$\cite{rehren1}.
For AAdS spacetimes as in Theorem \ref{t2}, however, it may happen that $\mathscr{Q}_{p,q}
\supsetneqq\mathscr{O}_{p,q}$. Likewise, defining $\mathscr{E}_{p,q}\doteq
\partial I^+(p,\widehat{\mathscr{M}})\cap\partial I^-(q,\widehat{\mathscr{M}})$,
let's start from

\begin{equation}\label{ads25}
\widetilde{\mathscr{Q}}_{p,q}=\bigcap_{r\in\mathscr{E}_{p,q}}\mathscr{W}_{s'_3(r),s_3(r)},
\end{equation}

where $s'_3(r)$ corresponds to $s_3(r)$ if we exchange future with past
in the statement of Proposition \ref{p3}. Here, $\widetilde{\mathscr{Q}}_{p,q}
\subset\mathscr{Q}_{p,q}$ is again causally complete, \emph{if nonvoid}. However, if
the spacetime metric deep inside the bulk is sufficiently ``distorted'', and
causing a sufficient number of null generators of, say, $\partial I^-(q,\widehat{
\mathscr{M}})$ to acquire a pair of conjugate points between $\mathscr{E}_{p,q}$ and
$\mathscr{I}$, for all we know (Proposition \ref{p3}) $\widetilde{\mathscr{Q}}_{p,q}$
could very well be empty (the intersection of the corresponding algebras may even
be nonvoid, but then we won't be able to attribute any localization whatsoever to
this algebra). This is suggested by the following remarks:

\begin{enumerate}
\item In a causally simple spacetime, any relatively compact diamond $\mathscr{O}_{p,q}$ is a
globally hyperbolic region, for which any Cauchy surface has a boundary equal to
$\mathscr{E}_{p,q}$;
\item Any causally complete region $\mathscr{U}$ has the following property: if
$\mathscr{S}\subset\mathscr{U}$ is a closed, achronal set with respect to $\mathscr{U}$, 
then $D(\mathscr{S})\subset\mathscr{U}$. 
\end{enumerate}

Both remarks together show that, if $r\in\mathscr{E}_{p,q}$ is such that a null generator
of, say, $\partial I^-(q,\widehat{\mathscr{M}})$ crossing $r$ acquires
a pair of conjugate points between $r$ and $s_1(r)$, then, by causal simplicity,
it follows that there's a neighbourhood of $q$ that is causally disjoint from $s_3(r)$, and, 
therefore, $I^-(s_3(r),\mathscr{M})$ cannot contain a Cauchy surface for $\mathscr{O}_{p,q}$.
Since, on the other hand, this doesn't exclude the possibility that 
$\widetilde{\mathscr{Q}}_{p,q}$ may contain points outside $\mathscr{O}_{p,q}$ either,
it's by no means clear whether the collections of $\mathscr{Q}_{p,q}$
and $\widetilde{\mathscr{Q}}_{p,q}$ give bases for the topology of $\widehat{\mathscr{M}}$
or not. \\

One way to circumvent these problems could be to restrict ourselves to sufficiently small
diamonds, such that none of the null generators of $\partial I^+(q,\mathscr{M})$ can
travel far enough beyond $q$ in order to develop a pair of conjugate points. But there is
a situation such that, no matter how small the extension, it'll always cease to be achronal:
it's when $q$ \emph{itself} is conjugate to $s_1(r)$. In this limiting case, $s_1(r)=s_2(r)$
but $s_2(r)\neq s_3(r)$.\\

We'll show now that the key out of these problems is to try to build a region similar 
to $\widetilde{\mathscr{Q}}_{p,q}$, but employing, instead of the points $s_3(r)$, 
$s'_3(r)$ for $r\in\mathscr{E}_{p,q}$, the points for which the problem, entailed by 
Proposition \ref{p3} and mentioned above, is, in a certain sense, ``minimized''. To perform this task, 
we'll start from a different viewpoint, which will also eventually show that the
critical situation in the preceding paragraph is ruled out by null geodesic completeness. 
First, notice that, using an argument similar to the one used in
\cite{hawkellis} and \cite{wald} to prove the existence of a (Lipschitz) topological manifold
structure for achronal boundaries, one can show that $\mathscr{E}_{p,q}$ is
locally the graph of a $\mathbb{R}$-valued, locally Lipschitz function of
$n-2$ real arguments, and therefore it's a compact, acausal, embedded (Lipschitz) 
topological submanifold of $\widehat{\mathscr{M}}$, with codimension 2. 
Notice as well that one can smoothly parametrize the family of timelike 
generators of $(\mathscr{I},b)$ by a latter's Cauchy surface $\mathscr{S}$ 
which is homeomorphic to $S^{n-2}$ and thus also compact. 
Let $t$ be the common affine parametrization of the timelike generators of 
$(\mathscr{I},b)$ mentioned in the previous Subsection. Define the function

\begin{equation}\label{ads26}
\tau:\mathscr{E}_{p,q}\times\mathscr{S}\ni(r,\theta)\mapsto\tau(r,\theta)\in\mathbb{R},
\end{equation}

where 

\begin{equation}\label{ads27}
\partial I^+(R,\mathscr{M})\cap T(\theta)=\{T(\theta)(\tau(r,\theta))\}.
\end{equation}

Proposition \ref{p2} shows that the definition of $\tau$ is not 
empty. Moreover:

\begin{proposition}
\label{p4}{\upshape\quad $\tau$ is upper semicontinuous in $r$ for fixed $\theta$.}
\begin{quote}{\small\scshape Proof.\quad}
{\small\upshape Let $\epsilon>0$. $r$ lies in the chronological past
of the point $T(\theta)(\tau(r,\theta)+\epsilon)$, and thus there's
an open neighborhood $U$ of $r$ in $\mathscr{E}_{p,q}$ which lies in 
the chronological past of $T(\theta)(\tau(r,\theta)+\epsilon)$. 
Therefore, for all $r'\in U$, we must have $\tau(r',\theta)<
\tau(r,\theta)+\epsilon$.\hfill$\Box$}\end{quote}
\end{proposition}

One can actually prove that $\tau$ is Lipschitz continuous in $\theta$
for fixed $r$, but this won't be used in the sequel. The function 
$\tau(.,\theta)$ will be called \emph{future Fermat potential}
with respect to $\theta$. The name is remnant of the Huygens-Fermat
principle of geometrical optics (see, for instance, pages 249-250
of \cite{arnold}). Now, extend the definition of $\tau(.,\theta)$
to the closure $\overline{\mathscr{F}_{p,q}}$ of some Cauchy surface 
$\mathscr{F}_{p,q}$ for $\mathscr{O}_{p,q}$, denoting it by the 
same symbol, since no confusion arises here. By the same argument
employed in Proposition \ref{p4}, $\tau(.,\theta)$ is upper
semicontinuous in $\overline{\mathscr{F}_{p,q}}$. Since both $\mathscr{E}_{p,q}
=\partial\mathscr{F}_{p,q}$ and $\overline{\mathscr{F}_{p,q}}$ are 
closed subsets of the compact set $\overline{\mathscr{O}_{p,q}}$,
they are compact themselves. By a standard result of analysis
(see, for instance, pages 110-111 of \cite{fomin}), $\tau(.,\theta)$
has a maximum value both in $\overline{\mathscr{F}_{p,q}}$ and
$\mathscr{E}_{p,q}$. The next theorem shows that $\tau(.,\theta)$
has indeed a distinguishing property of potentials:

\begin{theorem}
\label{t3}{\scshape(Maximum principle for the future Fermat potential) 
\upshape\quad The maximum value of $\tau(.,\theta)$ in
$\overline{\mathscr{F}_{p,q}}$ is achieved at $\mathscr{E}_{p,q}$.}
\begin{quote}{\small\scshape Proof.\quad}
{\small\upshape Let $r$ be a point of $\mathscr{E}_{p,q}$ where
$\tau(.,\theta)$ achieves its maximum in $\mathscr{E}_{p,q}$, and let 
$r'$ be a point of $\mathscr{F}_{p,q}$ such that $\tau(r',\theta)
\geq\tau(r',\theta)$. In such a case, it's obvious that $\mathscr{E}_{p,q}$
lies in the causal past of $T(\theta)(\tau(r',\theta))$. Pick a curve
segment in $\overline{\mathscr{F}_{p,q}}$ starting at $r'$, initially pointing
outside $J^-(T(\theta)(\tau(r',\theta)),\mathscr{M})$ and
ending in some point of $\mathscr{E}_{p,q}$. Then, any such a curve segment
must cross $\partial I^-(T(\theta)(\tau(r',\theta)),\mathscr{M})$
at least once more after $r'$, and before or at $\mathscr{E}_{p,q}$.
This shows that $\partial I^-(T(\theta)(\tau(r',\theta)),\mathscr{M})\cap
\overline{\mathscr{F}_{p,q}}$ encloses an open subset $X$ of 
$\mathscr{F}_{p,q}$ outside the causal past of $T(\theta)(\tau(r',\theta)$.\\

The remaining of the proof is analogous to the proof of Penrose's singularity
theorem\cite{hawkellis, wald}: namely, we'll show that the properties of $\partial X$
imply that there must exist an incomplete null geodesic in $(\widetilde{\mathscr{M}},
\widetilde{g}_{ab})$. First,
we'll show that the closed, acausal set $\partial X=\partial I^-(T(\theta)(\tau(r',\theta)),\mathscr{M})\cap
\overline{\mathscr{F}_{p,q}}$ is \emph{past trapped}, i.e., $\partial I^-(X,\widehat{\mathscr{M}})$
is \emph{compact}. The past ``ingoing'' null geodesics of $\partial X$ constitute
the past Cauchy horizon of $X$, which is thus contained in $\overline{\mathscr{O}_{p,q}}$
and therefore compact, as it's closed\cite{hawkellis,wald}. The past ``outgoing''
null geodesics are precisely the null generators of $\partial I^-(T(\theta)(\tau(r',\theta)),\mathscr{M})$
that cross $\partial X$. Given a common affine parametrization to the null generators
of $\partial I^-(T(\theta)(\tau(r',\theta)),\mathscr{M})$ such that the zero value of the 
affine parameter corresponds to $\partial X$. Then, let $t_0$ the largest value of
affine parameter for which a past endpoint of $\partial I^-(T(\theta)(\tau(r',
\theta)),\mathscr{M})$ is achieved. It must be finite, for each inextendible
null geodesic must acquire a pair of conjugate points before reaching infinity, 
although the value of the affine parameter at a past endpoint of the null generator 
segment starting at, say,  $r''\in\partial X$ can be zero if $r''$ happens to be itself a past endpoint.
Anyway, the portion of $\partial I^-(T(\theta)(\tau(r',\theta)),\mathscr{M})$
in the causal past of $\partial X$, being closed, has a closed inverse image in 
the compact set $[0,t_0]\times\partial X$ under the chosen parametrization of the null 
generators, and is therefore compact. Hence, the set $\partial I^-(\partial X,\widehat{\mathscr{M}})=
H^-(X)\cup\partial X\cup(\partial I^-(T(\theta)(\tau(r',\theta)),\mathscr{M})\cap J^-
(\partial X,\widehat{\mathscr{M}}))$ is a compact, achronal subset of $\widehat{\mathscr{M}}$, as asserted. \\

However, any causally simple spacetime is stably causal\cite{gr}. That is, one
can smoothly foliate $\widehat{\mathscr{M}}$ by ``constant-time'', spacelike surfaces of
codimension 1. By the structure of the conformal infinity, such surfaces (\emph{leaves}) cannot be 
compact. Moreover, each timelike \emph{orbit} of the foliation crosses an achronal set
at most once. By following these orbits, one can continuously map $\partial I^-(\partial 
X,\widehat{\mathscr{M}})$ into a spacelike leaf of this foliation. As the image of this map is compact, 
it must have a nonvoid boundary. But it's known that a set of the form $\partial I^-(Y,\widehat{\mathscr{M}}),
Y\subset\widehat{\mathscr{M}}$ is a topological submanifold without boundary of $\widehat{\mathscr{M}}$,
and, as such, it cannot have a boundary. This shows that some null generator of
$\partial I^-(T(\theta)(\tau(r',\theta)),\mathscr{M})$ must terminate at a singularity
before reaching its past endpoint. But this conflicts with the null geodesic completeness
of $\widehat{\mathscr{M}}$, entailed by asymptotic simplicity. Hence, no point in
$\mathscr{F}_{p,q}$ can achieve a maximum for $\tau(.,\theta)$ in
$\overline{\mathscr{F}_{p,q}}$ -- this maximum always takes place at $\mathscr{E}_{p,q}$.
\hfill$\Box$}\end{quote}
\end{theorem}

Propositions \ref{p4} and Theorem \ref{t3} together show that, 
for each $\theta$, there will always be a $r\in\mathscr{E}_{p,q}$ such that, given
\emph{any} Cauchy surface $\mathscr{F}_{p,q}$ for $\mathscr{O}_{p,q}$,
the set $\overline{\mathscr{F}_{p,q}}$ will always lie in the causal
past of $T(\theta)(\tau(r,\theta))$. By Proposition \ref{p3} and
the remarks above, this can only happen if the achronal null geodesic
segment $\gamma(r,\theta)$ linking $r$ to $T(\theta)(\tau(r,\theta))$ crosses $q$. 
Thus, this maximum point is \emph{unique:} suppose otherwise.
Then, there would be another $r'\in\mathscr{E}_{p,q}$ such that
there is an achronal null geodesic segment $\gamma(r',\theta)$ linking $r'$ to
$T(\theta)(\tau(r',\theta))=T(\theta)(\tau(r,\theta))$ and crossing $q$.
Now consider the curve segment $\gamma'(r,\theta)$ which coincides with
$\gamma(r,\theta)$ from $r$ to $q$, and coincides with $\gamma(r',\theta)$
from $q$ to $T(\theta)(\tau(r,\theta))$. This segment is necessarily broken,
which conflicts with the achronality of $\gamma(r,\theta)$. Exchanging the 
roles of $r$ and $r'$, one see that this argument also conflicts with
the achronality of $\gamma(r',\theta)$.
Notice, however, that an arbitrary $r\in\mathscr{E}_{p,q}$ \emph{need not} maximize
$\tau(.,\theta)$ for some $\theta$. Two instances where this cannot occur are:

\begin{enumerate}
\item $r$ is conjugate to $q$ along a null generator of $\partial I^-(q,
\widehat{\mathscr{M}})$ -- any future extension of this generator beyond $q$
won't be achronal;
\item $q$ is conjugate to $s_2(r)$ along a null generator of $\partial I^+(q,
\mathscr{M})$, by the remarks made above.
\end{enumerate}

The second instance, however, is excluded by our line of reasoning, because it renders impossible,
by Proposition \ref{p3} and Theorem \ref{t3}, to $\tau(.,\theta)$ to achieve
a maximum value in $\mathscr{E}_{p,q}$. This cannot happen, since for every
$\theta$ a maximum must exist by Proposition \ref{p4}. The first instance
can be circumvented by picking $\mathscr{O}_{p,q}$ contained, say, in
a convex normal neighbourhood, which can always be done, as here $(\widehat{\mathscr{M}},
\widehat{g})$ is strongly causal. One can go further and take
$\mathscr{O}_{p,q}$ sufficiently small (yet nonvoid) so that every
$r\in\mathscr{E}_{p,q}$ is a maximum point of $\tau(.,\theta)$ for some
$\theta$, as the only obstacle to this would be the second instance above, which 
is excluded by the above argument.
All results above have a past counterpart, by exchanging $q$ with $p$
and reversing the time orientation. \\

Summing up, we have showed that
\emph{sufficiently small $\mathscr{O}_{p,q}$ can always be precisely enveloped
by wedges,} by means of the prescription (\ref{ads22}). In such a case, any point not belonging to $\overline{\mathscr{O}_{p,q}}$
lies either in the chronological future of $\partial I^-(q,\widehat{\mathscr{M}})$
or in the chronological past of $\partial I^+(p,\widehat{\mathscr{M}})$,
and, as such, will fail to belong to some wedge enveloping $\mathscr{O}_{p,q}$.
Since the points at $\partial\mathscr{O}_{p,q}$ are already excluded from
the intersection by construction, one concludes that $\mathscr{O}_{p,q}=
\mathscr{Q}_{p,q}$ for suficiently small $\mathscr{O}_{p,q}$. Moreover, in
such a situation, each wedge in the definition (\ref{ads23}) of $\mathscr{Q}_{p,q}$
is guaranteed to be contained in some Poincaré domain. Therefore, one can even
restrict to a Poincaré domain and perform the bulk reconstruction there starting from a
boundary CFT in Minkowski spacetime.\\

\section{\label{sec4} Perspectives and open problems}

For \emph{additive} local quantum theories, it suffices to specify the
localization of the procedures for a basis of the manifold topology. Therefore,
the results in the previous Section indicate that one can completely recover
the \emph{bulk} quantum theory by just employing localization data from the \emph{boundary}
quantum theory and the Rehren bijection, and this theory is guaranteed to be causal if its holographic dual is.
In situations where the \emph{boundary theory} is additive, then all compactly
localized bulk observables are necessarily multiples of the identity\cite{rehren1}.
In such a case, one suffices to have just wedge localization in the bulk.\\

The covariance issue is obviously more complicated than in the AdS case. For
a proper implementation of conformal covariance in the boundary theory, two
diffeomorphisms which are ``asymptotic isometries''\cite{ashtekar} which differ
only by a difeomorphism which is an ``asymptotic identity'' (i.e., acting as the
identity on the boundary) should differ, from the viewpoint of the boundary theory,
only by an internal (non-geometric) symmetry. The lack of bulk isometry groups cries out
for a locally covariant formalism for local quantum physics, such as the one
developed in \cite{brunetti}. Algebraic holography then maps \emph{the realization of
a locally covariant} quantum theory in the bulk to a \emph{globally conformally covariant}
quantum theory at the boundary, where the latter has, in principle, an enormous amount of 
internal symmetry. For the conformal group to be unitarily implementable in some GNS
representation, these internal symmetries should not be generating a non-trivial
cohomological obstruction. If the state asssociated to the GNS representation 
satisfies the Reeh-Schlieder property, it follows from Proposition \ref{p1}, 
the discussion following it, and
the work of Brunetti, Guido and Longo\cite{longo} that, due to the breakdown of Haag duality,
the Tomita-Takesaki modular groups associated with the diamond von Neumann algebras
cannot unitarily implement the isotropy groups of the respective diamonds. From this,
it follows that either (or both) (1.) The unitary representation of the conformal group cannot be of
positive energy, or (2.) The conformal group is spontaneously broken. Both scenarios are of
the greatest interst for further study, as well as the possibility that such a spontaneous 
breaking has a cohomological structure stemming from the nontrivial asymptotic identities, 
and possible connections with the phenomenon of holographic Weyl anomalies\cite{sken1}.
This may even reveal an holographic encoding of bulk gravitational degrees of freedom into
the modular structure of the boundary theory.\\

All the reasoning in Subsection \ref{envelope} applies equally well if
one wants to rebuild the bulk localization using (sufficiently small) bulk 
\emph{regular diamonds}\cite{guido,ruzzi} instead of ordinary ones. This makes 
it also a good starting point for studying how the superselection sector 
structure is is holographically mapped between both theories. This problem
will be attacked in forthcoming work.\\

\section*{Acknowledgements}

I'm greatly indebted to Karl-Henning Rehren for his comments and suggestions
built over a critical reading of my former paper\cite{ribeiro2}, and to Jens Mund 
for sharing with me his insights on the intrinsic characterization of wedges in
curved spacetimes. I'd also like to thank the help from Sumati Surya and A. P. Balachandran
regarding the Subsection \ref{causal}, and Fernando Auil for numerous discussions.


\begin{thebibliography}{100}

\bibitem[Arno]{arnold} V.I. Arnold. \textsl{Mathematical Methods
of Classical Mechanics.} 2nd. ed., Springer-Verlag, 1989.

\bibitem[AshD]{ashdas} A. Ashtekar, S. Das. \textsl{Asymptotically 
Anti-de Sitter Space-times: Conserved Quantities.} Class.Quant.Grav. 
\textbf{17} (2000) L17-L30.

\bibitem[AsMa]{ashtekar} A. Ashtekar, A. Magnon. \textsl{Asymptotically
    Anti-de Sitter Space-times.} Class.Quant.Grav. \textbf{1} (1984)
    L39-L44.

\bibitem[BeEE]{gr} J.K. Beem, P.E. Ehrlich, K.L. Easley. \textsl{Global
  Lorentzian Geometry.} 2nd. ed., M. Dekker, 1996.

\bibitem[Bord]{borde} A. Borde. \textsl{Geodesic Focusing, Energy
    Conditions and Singularities.} Class.Quant.Grav. \textbf{4}
    (1987) 343-356.

\bibitem[Bous]{bousso} R. Bousso. \textsl{The Holographic Principle.}
    Rev.Mod.Phys. \textbf{74} (2002) 825-874.

\bibitem[BoRa]{boura} R. Bousso, L. Randall. \textsl{Holographic Domains 
of Anti-de Sitter Space.} JHEP \textbf{0204} (2002) 057.

\bibitem[BrFV]{brunetti} R. Brunetti, K. Fredenhagen, R. Verch.
    \textsl{The Generally Covariant Locality Principle -- A New 
    Paradigm for Local Quantum Physics.} Commun.Math.Phys. \textbf{237} (2003) 31-68.

\bibitem[BrGL]{longo} R. Brunetti, D. Guido, R. Longo. \textsl{Modular Structure 
and Duality in Conformal Quantum Field Theory.} Commun.Math.Phys. \textbf{156} (1993)
201-219.

\bibitem[FeRo]{roman} C.J. Fewster, T.A. Roman. \textsl{Null Energy Conditions
    in Quantum Field Theory.} Phys.Rev. \textbf{D67} (2003) 044003.

\bibitem[Fri1]{friedrich1} H. Friedrich. \textsl{Cauchy Problems for the
    Conformal Vacuum Field Equations in General Relativity.} Commun.Math.Phys.
    \textbf{91} (1983) 445-472.

\bibitem[Fri2]{friedrich2} H.Friedrich. \textsl{Existence and Structure
    of Past Asymptotically Simple Solutions of Einstein's Field Equations with
    Positive Cosmological Constant.} J.Geom.Phys. \textbf{3} (1986) 101-117.

\bibitem[Fri3]{friedrich3} H. Friedrich. \textsl{Einstein Equations
    and Conformal Structure: Existence of Anti-de Sitter-type
    Space-Times.} J.Geom.Phys. \textbf{17} (1995) 125-184.

\bibitem[Gal1]{galloway1} G.J. Galloway. \textsl{Maximum Principles 
for Null Hypersurfaces and Null Splitting Theorems.} Annales Henri 
Poincar\'e \textbf{1} (2000) 543-567.

\bibitem[Gal2]{galloway2} G.J. Galloway. \textsl{Some global results 
for asymptotically simple spacetimes.} In: Proceedings of the 
Workshop "The conformal structure of space-times: geometry, analysis, 
numerics", Tübingen, April 2001 (J. Frauendiener, H. Friedrich,
eds.). Lect.Notes Phys. \textbf{604} (2002) 51-60.

\bibitem[GaWa]{gao} S. Gao, R.M. Wald. \textsl{Theorems on 
  gravitational time delay and related issues.} Class.Quant.Grav. 
  \textbf{17} (2000) 4999-5008.

\bibitem[GLRV]{guido} D. Guido, R. Longo, J.E. Roberts, R. Verch.
\textsl{Charged sectors, spin and statistics in quantum field 
theory on curved spacetimes.} Rev.Math.Phys. 13 (2001) 125-198.

\bibitem[Haag]{haag} R. Haag. \textsl{Local Quantum Physics.} 2nd. ed.,
    Springer-Verlag, 1996.

\bibitem[HawE]{hawkellis} S.W. Hawking, G.F.R. Ellis. \textsl{The Large
    Scale Structure of Space-Time.} Cambridge, 1973.

\bibitem[HeSk]{sken1} M. Henningson, K. Skenderis. \textsl{Holography 
  and the Weyl Anomaly.} Fortsch.Phys. \textbf{48} (2000) 125-128.

\bibitem[KoFo]{fomin} A.N. Kolmogorov, S.V. Fomin. \textsl{Introductory
Real Analysis} (Translation by R.A. Silverman). Dover, 1975.

\bibitem[Mald]{malda} J. Maldacena. \textsl{The Large N Limit of
    Superconformal Field Theories and Supergravity.}
    Adv.Theor.Math.Phys. \textbf{2} (1998) 231-252.

\bibitem[ONei]{oneill} B. O'Neill. \textsl{Semi-Riemannian Geometry.}
Academic Press, 1983.

\bibitem[PeSW]{sorkin} R. Penrose, R.D. Sorkin, E. Woolgar. \textsl{
A Positive Mass Theorem Based on the Focusing and Retardation of Null 
Geodesics.} arXiv:gr-qc/9301015.

\bibitem[Reh1]{rehren1} K.-H. Rehren. \textsl{Algebraic Holography.}
    Annales Henri Poincar\'e \textbf{1} (2000) 607-623.

\bibitem[Reh2]{rehren2} K.-H. Rehren. Private communication.

\bibitem[Ribe]{ribeiro2} P.L. Ribeiro. \textsl{Renormalization Group
    Flow in Algebraic Holography.} In: Proceedings of the International 
    Conference on Renormalization Group and Anomalies in Gravity and 
    Cosmology (M. Asorey, I.L. Shapiro, eds.). Nucl.Phys.B -- 
    Proc.Suppl. \textbf{127} (2004) 193-195 (arXiv:hep-th/0306024).

\bibitem[Ruzz]{ruzzi} G. Ruzzi. \textsl{Punctured Haag duality in 
locally covariant quantum field theories.} arXiv:math-ph/0402027.

\bibitem[Sury]{surya} S. Surya. Private communication.

\bibitem[Wald]{wald} R.M. Wald. \textsl{General Relativity.} Chicago
  University Press, 1984.

\bibitem[Witt]{witt} E. Witten. \textsl{Anti de Sitter Space and
Holography}, Adv.Theor.Math.Phys. \textbf{2} (1998) 253-291.

\bibitem[Wool]{woolgar} E. Woolgar. \textsl{The Positivity of Energy
    for Asymptotically Anti-de Sitter Spacetimes.} Class.Quant.Grav. 
    \textbf{11} (1994) 1881-1900.

\end{thebibliography}
\end{document}